\documentclass[twocolumn,preprintnumbers,amsmath,amssymb,aip]{revtex4}

\usepackage{graphicx}
\usepackage{dcolumn}
\usepackage{bm}
\usepackage{color}
\usepackage{setspace}

\begin{document}


\title{Observation of Magnetopiezoelectric Effect in Antiferromagnetic Metal EuMnBi$_{2}$}

\author{Y. Shiomi$^{\, 1,2,3}$}
\author{H. Watanabe$^{\, 4}$}
\author{H. Masuda$^{\, 1}$}
\author{H. Takahashi$^{\, 1}$}
\author{Y. Yanase$^{\, 4}$}
\author{S. Ishiwata$^{\, 1,5}$}

\affiliation{$^{1}$
Department of Applied Physics and Quantum-Phase Electronics Center (QPEC), University of Tokyo, Hongo, Tokyo 113-8656, Japan}
\affiliation{$^{2}$
RIKEN Center for Emergent Matter Science (CEMS), Wako 351-0198, Japan}
\affiliation{$^{3}$
Department of Basic Science, University of Tokyo, Meguro, Tokyo 153-8902, Japan}
\affiliation{$^{4}$
Department of Physics, Kyoto University, Kyoto 606-8502, Japan}
\affiliation{$^{5}$
PRESTO, Japan Science and Technology Agency, Kawaguchi 332-0012, Japan}

\date{\today}

\begin{abstract}
We have experimentally studied a magnetopiezoelectric effect predicted recently for magnetic metals with low crystal symmetries. In EuMnBi$_2$ with antiferromagnetic Mn moments at $77$ K, dynamic displacements emerge along the $a$ direction upon application of ac electric fields in the $c$ direction, and increase in proportion to the applied electric fields. Such displacements are not observed along the $c$ direction of EuMnBi$_{2}$ or EuZnBi$_{2}$ with nonmagnetic Zn ions. As temperature increases from $77$ K, the displacement signals decrease and disappear at about $200$ K, above which electric conduction changes from coherent to incoherent. These results demonstrate the emergence of the magnetopiezoelectric effect in a magnetic metal lacking inversion and time-reversal symmetries.        
\end{abstract}

\maketitle

Piezoelectric effects that allow conversion between electricity and mechanical stresses are essential for a variety of applications, such as sensors, transducers, and actuators \cite{piezo-book, piezo-review}. To date, many examples of piezoelectric materials {\it e.g.} quartz \cite{Curie}, bones \cite{majid}, barium titanates \cite{Hippel}, lead zirconate titanates \cite{panda}, and alkaline niobates \cite{saito} have been found; all the materials are insulators. For metals, static electrical polarization is screened by conduction electrons and ceases to be well defined. The search for piezoelecric materials thereby has been carried out with insulators and semiconductors. Nevertheless, the realization of highly-conductive piezoelectric materials could offer new applications of piezoelectricity because of their compatibility with electrical circuits \cite{shiomi-AEM}. 
\par

Magnetopiezoelectric effect (MPE), which was theoretically predicted very recently \cite{watanabe, MPE-PRL, watanabe-arxiv}, has a potential to realize piezoelectric {\it metals}. For magnetic metals that simultaneously break time-reversal and space-inversion symmetries, the breaking of the symmetries results in dynamical distortion in response to ac electric currents \cite{watanabe, MPE-PRL, watanabe-arxiv}. Electric currents induce an electronic nematic order in metals lacking inversion and time-reversal symmetries \cite{watanabe}; the nematic order accompanies the modulation of Fermi surfaces, which in turn leads to a structural deformation through electron-lattice couplings. The MPE thereby requires Fermi surfaces, and interestingly, is expected to be free from screening effects unlike conventional piezoelectric effects \cite{watanabe, watanabe-arxiv}. Also, the MPE can be viewed as a generalization of magnetoelectric responses in insulators, and its relation to Berry phase effects has been discussed \cite{MPE-PRL}. 
\par

Dynamic strains caused by ac electric currents in the MPE is apparently a metallic analog of the inverse piezoelectric effect, but the MPE differs from the conventional piezoelectric effect in terms of symmetry \cite{watanabe, MPE-PRL, watanabe-arxiv}. In the MPE, both time-reversal and space-inversion symmetries need to be broken, and hence the MPE occurs only in magnetically ordered states. In contrast, the conventional piezoelectric effect respects time-reversal symmetry. From the symmetry argument, the conventional piezoelectricity does not arise in magnetopiezoelectric metals where magnetic order breaks space inversion symmetry \cite{watanabe}.   
\par

\begin{figure}[t]
\begin{center}
\includegraphics[width=8.5cm]{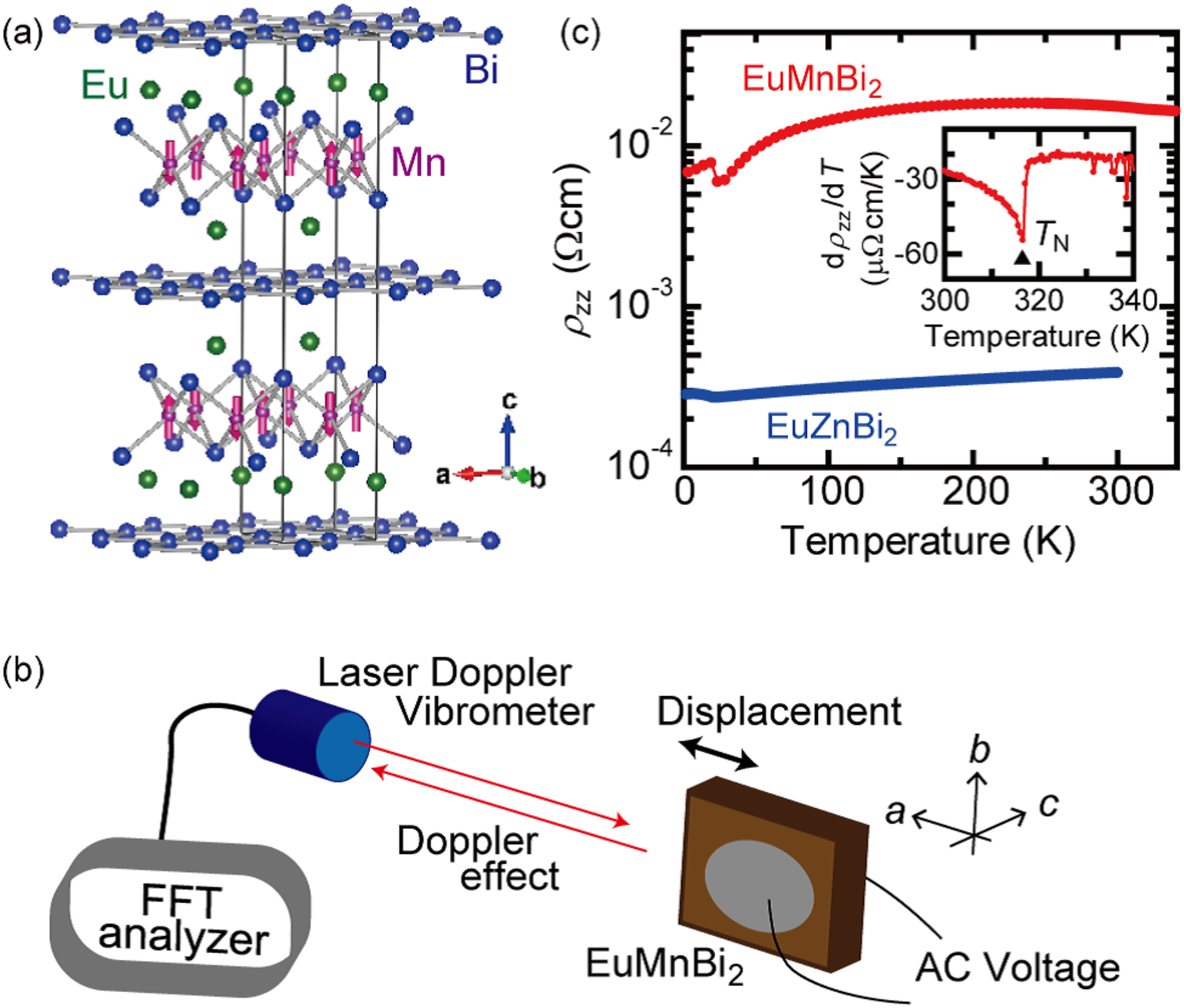}
\caption{(Color online.) (a) Schematic illustration of crystal structure for EuMnBi$_{2}$, together with the magnetic structure of Mn ions below the antiferromagnetic transition temperature ($T_{N}$) [\onlinecite{guo, masuda-neutron}]. (b) Schematic illustration of measurement setup for the magnetopiezoelectric effect. (c) Temperature dependence of interlayer resistivity $\rho_{zz}$ for EuMnBi$_{2}$ and EuZnBi$_{2}$. The inset shows temperature derivative of resistivity for EuMnBi$_{2}$ around $T_{N}(=315$ K).  } 
\label{fig1}
\end{center}
\end{figure}

In this letter, we experimentally study the MPE for antiferromagnetic metal EuMnBi$_{2}$ \cite{may, masuda, masuda-prb}. In the antiferromagnetic state, EuMnBi$_{2}$ has the same crystal symmetry as hole-doped BaMn$_{2}$As$_{2}$, the MPE of which was discussed theoretically \cite{watanabe}. As shown in Fig. \ref{fig1}(a), antiferromagnetic EuMnBi$_{2}$ possesses the $D_{2d}$ symmetry \cite{guo, masuda-neutron}, which allows the magnetopiezoeleoctric response given by $\varepsilon_{xy}=e_{zxy}E_{z}$ \cite{watanabe}. Namely, an electric field along the $z$ axis ($E_{z}$) induces a stress along the [110] direction ($\varepsilon_{xy}$) which corresponds to a tetragonal-to-orthorhombic structural deformation \cite{kasahara}. Here, $e_{zxy}$ is a magnetopiezoelectric coefficient, a measure of the piezoelectric conversion efficiency. 
\par

We employ laser Doppler vibrometry \cite{shiomi-AEM, herdier, mccartney, shetty} in a low-temperature environment to detect very small magnetopiezoelectric signals. The laser Doppler vibrometry enables non-contact detection of sub-pm-level vibrations even for thin films \cite{mccartney, shetty} and microstructures \cite{seo}, by measuring the Doppler frequency shift of the scattered light from the sample using a two-beam laser interferometer [Fig. \ref{fig1}(b)]. By applying ac electric fields along the $c$ axis of EuMnBi$_{2}$, we have observed dynamic displacements satisfying the relation $\varepsilon_{xy}=e_{zxy}E_{z}$ at $77$ K. Such displacements are not observed for a paramagnetic analogue EuZnBi$_{2}$. The dynamic displacements observed in EuMnBi$_{2}$ decreases with increasing temperature, and becomes undetectable levels above $\sim 200$ K, the crossover temperature of coherent to incoherent (hopping-like) conduction. The present results provide experimental evidence of the MPE, and also reveal that the laser Doppler vibrometry is a useful probe of the nematicity which has been reported {\it e.g.} in cuprate superconductors \cite{kohsaka, lawler, hinkov, ando, daou, damascelli}, iron-based superconductors \cite{kasahara, chu}, and heavy-electron systems \cite{okazaki, tonegawa, ikeda, matsubayashi}.
\par

Single crystals of EuMnBi$_{2}$ and EuZnBi$_{2}$ were grown by a Bi-flux method according to the papers reported by some of the present authors \cite{masuda, masuda-prb}. Plate-like single crystals of EuMnBi$_{2}$ with the size of $3 \times 3 \times 1$ mm$^{3}$ were fixed to a copper sample holder using GE varnish. On the largest planes corresponding to the $c$ planes, voltage electrodes were formed using conductive silver pastes, as illustrated in Fig. \ref{fig1}(b). While applying kHz-range ac voltages to the samples along the $c$ direction, time-dependent displacements generating along the $a$ direction were measured using a laser Doppler vibrometer combined with a fast Fourier transform (FFT) analyzer \cite{shiomi-AEM, herdier, mccartney, shetty} [Fig. \ref{fig1}(b)]. A red laser is directed at the surface of EuMnBi$_{2}$, and the vibration velocity of the sample is extracted from the Doppler shift of the reflected laser. The observed velocity was then numerically integrated with respect to time using the FFT analyzer, to obtain the vibration amplitude of the sample. With an objective lens, the laser spot diameter is set to be less than $100$ micron, which can be smaller than domain sizes of bulk antiferromagnets \cite{fiebig}. Low temperature experiments were conducted with a nitrogen optistat having a quarts window. 
\par

First, we show temperature dependence of interlayer resistivity $\rho_{zz}$ for EuMnBi$_{2}$ and EuZnBi$_{2}$ in Fig. \ref{fig1}(c). As temperature decreases from $340$ K, $\rho_{zz}$ of EuMnBi$_{2}$ exhibits a broad maximum at about $200$ K, and decreases gently. Such a crossover from non-metallic ($d\rho_{zz}/dT <0$) to metallic ($d\rho_{zz}/dT >0$) interlayer transport has been reported in anisotropic layered metals \cite{yoshida, valla, gutman, terasaki, loureiro, tsukada}. Because of anisotropic Fermi surfaces, hopping-type conduction across blocking layers is dominant at high temperatures, whereas at low temperatures, the system behaves as an anisotropic three-dimensional metal and the transport becomes metallic. In the coherent transport regime, the conductivity is determined by a scattering time (quasiparticle lifetime) and the Fermi velocity \cite{valla}, but such a band picture is not valid in the incoherent regime. In contrast to the case of EuMnBi$_{2}$, $\rho_{zz}$ of EuZnBi$_{2}$ is metallic ($d\rho_{zz}/dT >0$) in the entire temperature regime in Fig. \ref{fig1}(c), which reflects weaker anisotropy of EuZnBi$_{2}$ than EuMnBi$_{2}$. At approximately $20$ K, $\rho_{zz}$ for EuMnBi$_{2}$ and EuZnBi$_{2}$ shows an anomaly, corresponding to magnetic transition of Eu ions \cite{masuda, masuda-prb}. As shown in the inset to Fig. \ref{fig1}(c), the antiferromagnetic transition temperature of Mn ions in EuMnBi$_{2}$ is determined to be $315$ K \cite{may} from the temperature derivative of resistivity. 
\par

\begin{figure}[t]
\begin{center}
\includegraphics[width=8.5cm]{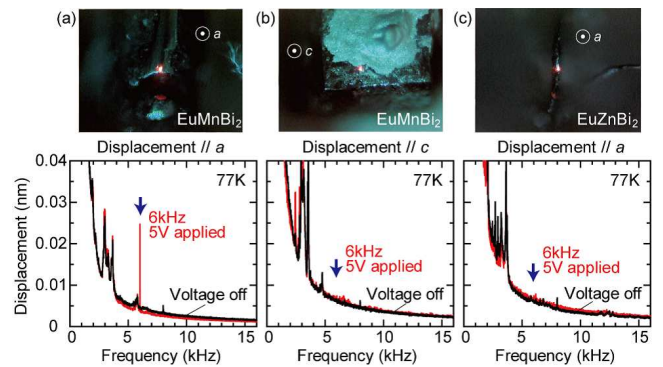}
\caption{(Color online.) Frequency dependence of displacement signals measured at $77$ K with (red color) and without (black color) voltage application to (a),(b) EuMnBi$_{2}$ and (c) EuZnBi$_{2}$ samples. Here, the voltage of $6$-kHz frequency and $5$-V amplitude ($10$-V peak-to-peak amplitude) was applied along the $c$ direction of the samples. The displacement along the $a$ direction was measured for (a) EuMnBi$_{2}$ and (c) EuZnBi$_{2}$, while that along the $c$ direction was measured for EuMnBi$_{2}$ in (b). In the top panels, sample pictures are also shown; red points correspond to the laser spots.   } 
\label{fig2}
\end{center}
\end{figure}

We then performed measurements of the MPE for EuMnBi$_{2}$ at $77$ K. Figure \ref{fig2} shows FFT spectra of displacements for EuMnBi$_{2}$ and EuZnBi$_{2}$ with and without application of voltage. In Fig. \ref{fig2}(a), the displacement was measured along the $a$ direction of EuMnBi$_{2}$ while applying voltage of $6$-kHz frequency to the $c$ direction. When the voltage is off, no signals are observed at the frequency of the voltage ($6$ kHz); note that at low frequencies, the displacement signal diverges because the numerical integration of velocity data in the frequency domain corresponds to division of the velocity by frequency. However, when $5$-V voltage of $6$-kHz frequency is applied, a peak signal is clearly observed at the voltage frequency. The magnitude of the displacement at $6$ kHz is $0.025$ nm. See also Supplemental Material (SM) for the laser-position dependence and the voltage-frequency dependence of the displacement signal. Since heating effects ($\propto E_{z}^{2}$) should give rise to second-harmonic signals (at $12$ kHz), the clear displacement signal at the voltage frequency agrees with the MPE.  
\par

To check the consistency with the MPE, we measured displacements along the $c$ direction of the same EuMnBi$_{2}$ sample while applying ac voltage in the $c$ direction in Fig. \ref{fig2}(b). When the laser direction is parallel to the voltage direction, we found that the displacement signal at the voltage frequency is smaller than the measurement limit even in the case of application of $5$-V voltage. This direction dependence is consistent with the MPE expected from the symmetry in EuMnBi$_{2}$ \cite{watanabe}; for ac voltage applied to the $c$ direction, the distortion is expected along the [110] direction, but not along the $c$ direction. We also confirmed that displacement signals are not observed in response to applied voltages for isostructural paramagnet EuZnBi$_{2}$, as shown in Fig. \ref{fig2}(c). Antiferromagnetic Mn ions are therefore necessary for the emergence of the displacement signals.  
\par

\begin{figure}[t]
\begin{center}
\includegraphics[width=8.5cm]{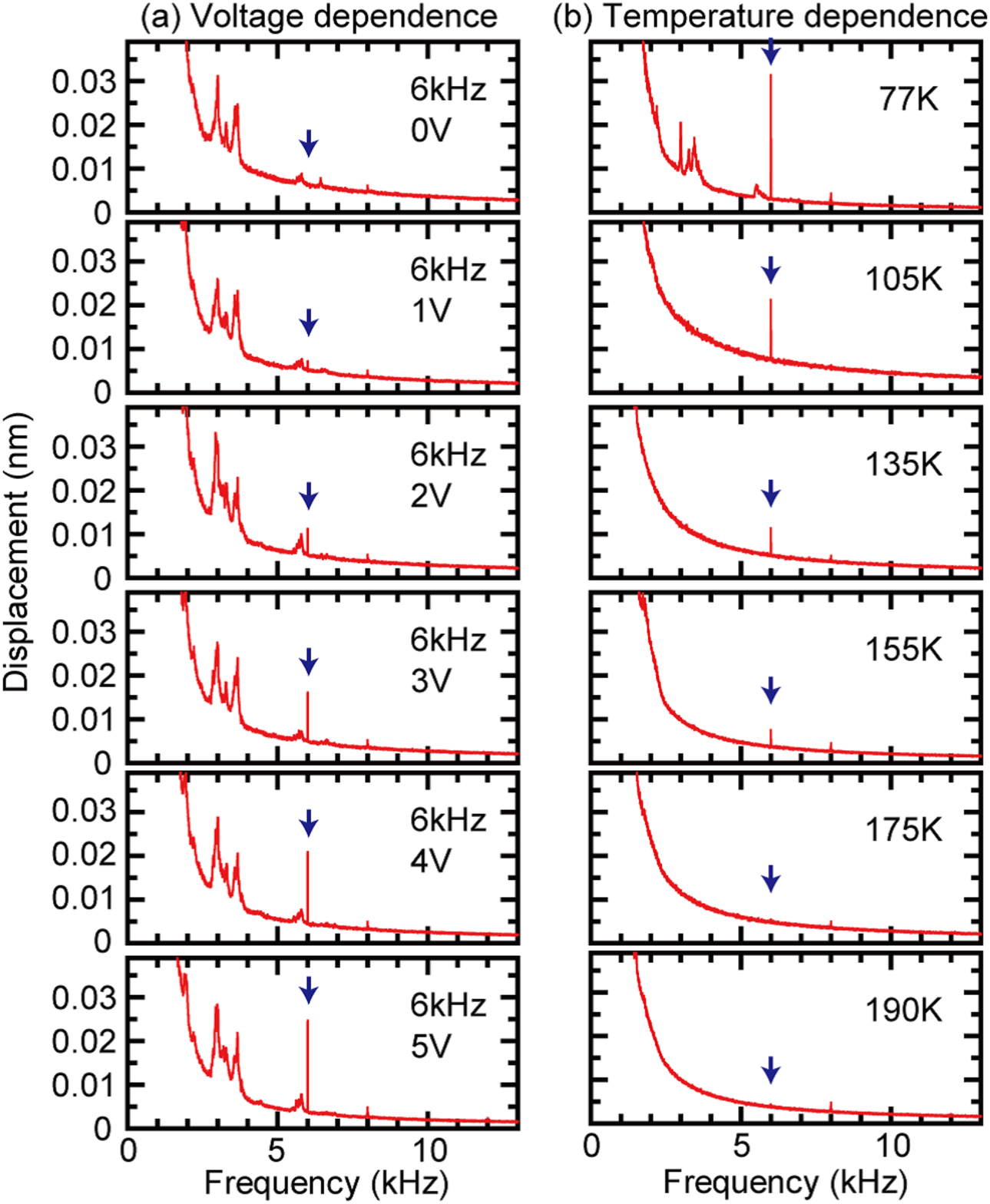}
\caption{(Color online.) (a) Frequency dependence of the displacement signals measured with various voltage amplitudes for EuMnBi$_{2}$ at $77$ K. Here, the displacements were measured along the $a$ direction while the voltage of $6$-kHz frequency was applied to the $c$ direction. (b) Frequency dependence of the displacement signals measured for EuMnBi$_{2}$ at various temperatures. Here, the displacements were measured along the $a$ direction while the voltage of $6$-kHz frequency and $5$-V amplitude was applied to the $c$ direction. } 
\label{fig3}
\end{center}
\end{figure}

\begin{figure}[t]
\begin{center}
\includegraphics[width=8.5cm]{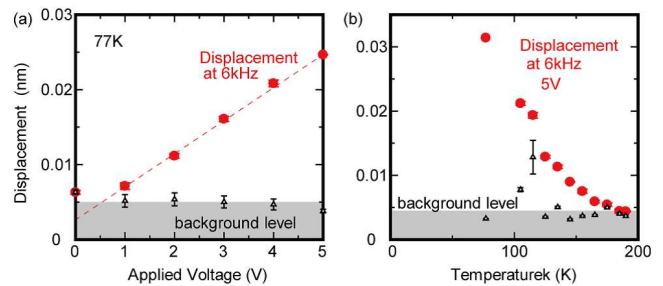}
\caption{(Color online.) (a) Voltage dependence of the displacement signal measured at $6$ kHz (red circles). See Fig. \ref{fig3}(a) for raw data. The background levels (black triangles) correspond to the averaged displacements around $6$ kHz without voltage application. The dotted line is guide for eyes. (b) Temperature dependence of the displacement signal measured at $6$ kHz (red circles). See Fig. \ref{fig3}(b) for raw data. The background levels (black triangles) correspond to the averaged displacements around $6$ kHz without voltage application. } 
\label{fig4}
\end{center}
\end{figure}

Voltage-amplitude dependence of the displacement signal is examined at $77$ K in Fig. \ref{fig3}(a). Here, the displacement is measured in the $a$ direction of EuMnBi$_{2}$ while the amplitude of $6$-kHz voltage is changed from $0$ V through $5$ V. At $0$ V, no signal is recognized, but the signal at $6$ kHz gradually increases with increasing voltage amplitudes. For $5$-V voltage corresponding to $\approx 100$ mA, the displacement reaches $0.025$ nm. The $6$-kHz displacement is plotted as a function of the voltage amplitude in Fig. \ref{fig4}(a). Obviously, the displacement at the voltage frequency increases in proportion to voltage intensity, consistent with the MPE. From a linear fit to the experimental data [dotted line in Fig. \ref{fig4}(a)], we noticed that there is a small offset independent of voltage application. This offset, whose magnitude is as small as $3$ pm, is ascribable to experimental noises, since the measured displacement signals should include noise signals from the measurement environment in addition to the MPE. 
\par

Temperature dependence of the MPE is also informative, because the magnitude of the MPE signal depends on transport, mechanical, and magnetic properties. The displacement signal at the $6$-kHz voltage frequency was thereby investigated at different temperatures in Fig. \ref{fig3}(b), although the data might include ambiguity because background levels inevitably change by temperature shift of the laser position. As temperature increases from $77$ K, the displacement signal at $6$ kHz decreases monotonically, and finally disappears at $175$ K. As shown in SM, we confirmed that the signal also disappears at approximately $200$ K for different two EuMnBi$_{2}$ samples. The onset temperature of $\sim 200$ K is much smaller than the antiferromagnetic transition temperature of EuMnBi$_{2}$ ($315$ K), where the magnetic order breaks space inversion symmetry. Instead, the emergence of the MPE seems to be related with the crossover of incoherent to coherent conduction [Fig. \ref{fig1}(b)]. In the high-temperature incoherent regime, the interlayer transport is no longer dictated by the Fermi-liquid picture, and the MPE becomes indiscernible.
\par

In Fig. \ref{fig4}(b), temperature dependence of the displacement signal observed at the voltage frequency is summarized. With increasing temperature from $77$ K, the displacement signal rapidly decreases almost as the inverse square of temperature. As mentioned before, the temperature dependence of the MPE signal depends on transport, mechanical, and also magnetic properties of materials, but has not been discussed in detail in the theoretical papers \cite{watanabe, MPE-PRL, watanabe-arxiv}. In fact, though the susceptibility of the current-induced nematic order was calculated to be proportional to scattering time \cite{watanabe}, the temperature dependence of the displacement is apparently greater than that of interlayer conductivity [Fig. \ref{fig1}(c)]. Owing to the strong anisotropy in Fermi surfaces, interlayer conductivity may not be proportional to scattering time \cite{gutman}. Also, temperature change in electron-lattice couplings affects that of the MPE signal. Furthermore, the MPE signal in EuMnBi$_{2}$ depends also on exchange couplings between itinerant Bi-electron spins and localized Mn moments, since conduction is governed by Bi-band electrons but magnetism originates from Mn moments. The spin exchange couplings can be stronger at lower temperatures. Though the authors in the preceding theoretical papers \cite{watanabe, watanabe-arxiv} discuss the MPE for antiferromagnetic metals from the symmetry argument, a microscopic model in which material parameters are taken into account will be necessary to elucidate the temperature dependence of the MPE signal in EuMnBi$_{2}$.     
\par

Finally, from the experimental results at $77$ K, we estimate the magnitude of magnetopiezoelectric coefficient $e_{zxy}$ for EuMnBi$_{2}$. Using the applied voltage $5$ V, the observed displacement $0.025$ nm, and the sample size, $e_{zxy}$ is calculated to be approximately $1$ pC/N. This magnitude is $1000$ times less than the piezoelectric constant for lead zirconate titanates \cite{panda}, and comparable to that of quarts \cite{Curie, You}. The piezoelectric performance of EuMnBi$_{2}$ is thereby insufficient for device applications, but one can expect that materials with higher conductivity (scattering time) have higher magnetopiezoelectric coefficients \cite{watanabe}. From this perspective, anisotropic EuMnBi$_{2}$ ($\rho_{zz}\approx 1 \times 10^{-2}$ ${\rm \Omega cm}$) is not suitable for large magnetopiezoelectric coefficients, and three-dimensional systems with isotropic Fermi surfaces {\it e.g.} Mn$_{2}$Au \cite{watanabe-arxiv} ($\rho\approx 1 \times 10^{-5}$ ${\rm \Omega cm}$ \cite{yu}) may be a better choice of materials. As far as only transport properties are considered, magnetopiezoelectric metals with high conductivity and high magnetic transition temperature could exhibit piezoelectric performance comparable to conventional piezoelectric materials even at room temperature.   
\par

In summary, the MPE has been demonstrated for antiferromagnetic metal EuMnBi$_{2}$ using laser Doppler vibrometry that can detect tiny sample vibrations sensitively. In response to applied ac voltages in the $c$ direction, an inplane dynamic displacement whose magnitude increases in proportion to applied voltage was observed for EuMnBi$_{2}$ single crystals at $77$ K, whereas not for paramagnetic relative EuZnBi$_{2}$. As temperature increases, the displacement signal due to voltage application decreases monotonically and disappears at approximately $200$ K.  This result indicates that the transport governed by coherent quasiparticles is necessary for the generation of the MPE signal. The MPE, a piezoelectric effect in magnetic metals, could lead to a new function in piezoelectric devices, as well as spintronic devices.   
\par


The authors thank Y. Tokura, N. Nagaosa, T. Nakajima, and K. Harii for fruitful discussions. This work was supported by JSPS (KAKENHI No. 15H05884, No. 16K17736, No. 17H01195, No. 17H04806, No. JP18H04225, No. JP18H04215, No. 18H01178, and No. 18H04311). Y.S. was supported by JST ERATO Grant Number JPMJER1402, Japan, and H.W. and H.M. were supported by JSPS through a research fellowship for young scientists (No. JP18J23115 and No. JP16J10114, respectively).  
\par

%




\end{document}